\documentclass[aps,preprint,showpacs]{revtex4-1}
\usepackage{amsmath, gensymb}
\usepackage{epsfig}
\usepackage{graphics}
\begin{document}

%\draft

\title{Model for domain wall avalanches in ferromagnetic thin films}
\author{R. C. Buceta}
\email{rbuceta@mdp.edu.ar}
\author{D. Muraca}
\altaffiliation[Present address: ]
{Instituto de Fisica Gleb Wataghin, Universidade Estadual de Campinas (UNICAMP)
Campinas, 13.083-970, SP, Brazil}
\affiliation{Instituto de Investigaciones F\'{\i}sicas de Mar del Plata, Universidad Nacional de Mar del Plata and Consejo Nacional de Investigaciones Cient\'{\i}ficas y T\'ecnicas, Funes 3350, B7602AYL Mar del Plata, Argentina}

\begin{abstract}
The Barkhausen jumps or avalanches in magnetic domain-walls motion between succesive pinned configurations, due the competition among magnetic external driving force and substrum quenched disorder, appear in bulk materials and thin films. 
We introduce a model based in rules for the domain wall evolution of ferromagnetic media with  exchange or short-range interactions, that include disorder and driving force effects. We simulate in 2-dimensions with Monte Carlo dynamics, calculate numerically distributions of sizes and durations of the jumps and find power-law critical behavior.
The avalanche-size exponent is in excellent agreement with experimental results for thin films and is close to predictions of the other models, such as like random-field and random-bond disorder, or functional renormalization group. The model allows us to review current issues in the study of avalanches motion of the magnetic domain walls in thin films with ferromagnetic interactions and opens a new approach to describe these materials with dipolar or long-range interactions.
\!\!\footnote{This version includes revisions.\vspace{1ex}} 
\end{abstract}
\pacs{64.60.av, 45.70.Cc}
\maketitle

\section{\label{sec:intro}Introduction}

Depending on shape, composition, and energy contributions, among others, ferromagnetic materials can form magnetic domains. These domains are separated by domain walls (DW) and their magnetization is slightly inhomogeneous, due to internal disorder. The disorder includes vacancies, defects, impurities and dislocations, fixed inside the material, and limits the movement of DW when the material is placed in an external magnetic field. The quenched disorder, known as Barkhausen noise by the name of the scientist who first observed,\cite{*[{}] [{. See a traslation from German in ref. \onlinecite{Durin-05}, pp. 253--257}]Barkhausen-19} is an early example of the origen of complexity in materials science and has been intensively studied to date.\cite{Urbach-95, Dahmen-94, *Perkovic-95,*Dahmen-00,*Carpenter-01,*Dahmen-01,*Mehta-02} In many cases, the wall exhibit a jerky motion between states of quiescence, known as Barkhausen jumps or avalanches.
The competition between restoring forces and quenched disorder results in transitory multi-metastable states. Locally, the wall leaves a metastable state basically by two mechanisms: by fluctuations spontaneously or, under certain conditions, driven by magnetic external force. Examples of avalanches with quenched sustrate disorder are the dislocation and crack propagation,\cite{Moretti-04,*Ponson-06,*Ponson-07,*Bonamy-08} stick-slip motion of tectonic plates\cite{Fisher-97,*Fisher-98,*Schwarz-03,*Jagla-10} and those with origin from depinning of fluid contact-interface.\cite{Prevost-02,*Prevost-99,*Moulinet-02,*Moulinet-04} Avalanches have also been studied in models without quenched substrate disorder, such as in sandpile models and in granular matter.\cite{Bak-87,*Tang-88,*Dhar-89,*Banerjee-95,*Dhar-06} An important characteristics of avalanche motion is its scale invariance, self-organized criticality, a distribution $P(S)\sim S^{-\tau}$ of the avalanche-sizes $S$, and a distribution $P(T)\sim T^{-\alpha}$ of the avalanches-durations $T$, being $\tau$ and $\alpha$ the size and duration critical exponents, respectivelly. 

The structure and motion of the DW in bulk materials with defects has been studied long time.\cite{Bertotti-98} In such systems the more accepted structure is that of Bloch, where the magnetization rotates through of wall, outside of the plane of the magnetic moments. The statistical properties of the DW avalanches are understood today by means of depinning transition models,\cite{Durin-05, Sethna-05x} that explain very well many experimental results. Models and experiments have been grouped into two universality classes depending on the values of critical exponents, which also verified scaling equalities fully accepted.\cite{Colaiori-08} 
One class includes materials with dipolar or long-range interactions, {\sl e.g.} polycrystaline and crystallized amorphous materials, where the exponents take values close to $\tau_{\ell}=1.5$ and $\alpha_{\ell}=2.0$\,.\cite{Durin-00}
The other class includes materials with predominantly exchange or short-range interactions, {\sl e.g.} soft-magnetic materials with high anisotropy, where the exponents take values close to $\tau_{s}=1.27$ and $\alpha_{s}=1.50$\,.\cite{Urbach-95, Bahiana-99, Durin-99}
When dipolar interactions are neglegect front propagation occurs, similarly {\sl e.g.} fluid imbibition in porous media, otherwise there is a mean field like behavior.
%$\tau_{s}=1.27$ {\pm 0.03} and $\alpha_{s}=1.5\pm 0.05$
%$\tau_{\ell}=1.5\pm 0.1$ and $\alpha_{\ell}=2.0\pm 0.2$

In thin films the wall motion is dominated by depinning too, but additionally the DW change their structure depending on the thickness of the film. Neel was the first to note that the theory of the Bloch walls is not valid for thin films if the thickness is comparable to the width of the wall.\cite{Neel-55,*Huber-58} As the thickness decreases, the wall passes through the stages of Bloch wall, asymmetric walls, crosstie wall, and N\'eel wall where atomic moments remain in plane while the rotation occurs.\cite{Hubert-98} It is not known yet if the wall structure affects movement. Some experimental results show that no changes expected in the statistical results when varying the thickness of the thin film.\cite{Santi-06} However, theoretically, the manner of interaction of the wall with the defects could determine whether the magnetization can rotate locally overcome the defect or, in other words, if the wall can pass across the defect.\cite{*[{Barkhausen avalanches in Co thin films have been observed directly by }] [{}] Kim-03a} The thin film exponents should be universal in the sense of describing a large number of materials and systems and must be independent of microscopic details. Magnetic thin films (bellow 200 nm) have exponents belonging to a new universality classes different from the 3-dimensional magnets.\cite{Puppin-00, Kim-03b, Ryu-07} The usual theoretical approach to these thin films are 2-dimensional models. Recently, experimental results show that the critical exponent of the size distribution decreases with increasing temperature, a fact attributed to a transition between the universality classes of long- to short-range interactions.\cite{Ryu-07} However, theoretically, this is an argument that can be sustained only for 3-dimensional systems.\cite{Sethna-07} 

The paper is organized as follows. In the next Section we present the simulation model and the statistical properties of the DW avalanches.  We introduce a model based on local evolution rules of the wall, with or without external magnetic field, taking into account the defects found the wall in your progress. Also we present the definitions of avalanche size and duration used here to calculate the distributions. In Section III we provide an overview of our numerical results. We present the basic statistical properties of the distributions of avalanche\--size and -duration. Also we analyze the results compared with available experimental and theoretical. Section IV finishes the paper with conclusions.

\section{\label{sec:model}The model}

In a magnetizable material we consider a volume element consisting of two domains separated by a magnetic DW. We assume both domains composed of opposite magnetic dipoles and point defects arranged in the nodes or sites on a regular array. We call active (or inactive) to sites containing dipoles (or defects). We suppose that the DW consists only of active sites and can move through the inactive ones.  In the model we disregard the possible mechanisms that explain the structure of the DW. Just assume that at any moment the DW is a monolayer dipole with magnetic dipole moment perpendicular to the domains, and that the magnetization is in the ``easy direction''. 

To model a thin film,  we consider dipoles and defects located at the nodes of a two-dimensional square lattice of edge $L$ with periodic boundary condition. 
In this lattice, we assign a random pinning function $\xi(i,k)$ uniformly distributed in the interval ͓$[0,1͔]$ to every node $(i,k)$ of the square lattice. For a given parameter $p\in[0,1]$, if $\xi(i,k)\le p$ the node $(i,k)$ is active, otherwise is inactive. To characterize the disorder in the lattice we use the activity function $F(i,k)=\theta(p-\xi(i,k))$,\cite{Braunstein-00, *Costanza-03} where $\theta(x)$ is the unit step function defined as $\theta(x)=1$ for $x\ge 0$ and $\theta(x)= 0$ for $x<0$. A dipole located at the nodes of the DW has perpendicular component to the easy direction. If $(j,n_j)$ is a point on the DW in the discrete lattice, with $j=1,\dots,L\,$ and $n_j=1,2,\dots$, an active site (or inactive) next to the wall is described by $F(j,n_j\pm 1)=1\;(\mathrm{or}\;0)$.

Our model is based on local evolution rules for all points of the DW. These rules include the short-range ferromagnetic interactions, taking into account the local balance of magnetic moment on both sides of a point on the wall. When there is local unbalance in the opposite direction of movement, the DW at this point spontaneously reaches equilibrium. Otherwise, when there is balance, the DW can move driven by the applied field. Hence we assume that a point on the DW can move spontaneously with probability $c$ if the magnetic dipole moment in the environment is unbalanced on both sides of the wall. Else, if there is an approximate balance on both sides, with probability $1\!-c\;$ the DW point on the wall is driven by action of the field that attempts to remove it from the metastable state. We start the simulation with a flat DW. Chosen column $j$, the point $(j,n_j)$ from the wall evolves according the following rules. 
\begin{itemize}
\item 
With probability $c\,$: if $F(j,n_j+1)=1$ and (a) $\,\tfrac{1}{2}(n_{j+1}+n_{j-1})-n_j\ge 1$ the wall point moves one unit [{\sl i.e.} $n_j\to n_j+1$], or (b) otherwise the wall point is pinned. 
\item
Else, with probability $1\!-\!c\,$: 
\begin{itemize}
\item If $F(j,n_j+1)\!=\!1$ and $n_{j+1}=n_{j-1}\ge n_j$ the wall point moves one unit up from its neighbors [{\sl i.e.} $n_j\to n_{j+1}+1$], otherwise the wall point moves (or not) up to the maximum between the site or its neighbors [{\sl i.e.} $n_j\to \max(n_{j+1},n_{j-1},n_{j})$]; 
\item If $F(j,n_j+1)\!=\!0$ and (a) $\,n_{j+1}=n_{j-1}> n_j+1$ the wall point moves one unit up from its neighbors [{\sl i.e.} $n_j\to n_{j+1}+1$], or (b) $\;n_{j+1}\neq n_{j-1}$ and $\max(n_{j+1},n_{j-1})> n_j+2$ the wall point moves up to the maximum between its neighbors [{\sl i.e.} $n_j\to  \max(n_{j+1},n_{j-1})$], or (c) otherwise the wall point is pinned. 
\end{itemize}
\end{itemize}
As the DW is made up of active sites, a rule is frustrated when a wall point trying to reach inactive sites.  

If $\mathrm{n}(t)=(n_1(t),\dots,n_L(t))$ is the position of the DW at time $t$, it is in metastable state during a time interval $\Delta t>0$ if $\mathrm{n}(t) = \mathrm{n}(t+\delta t)$ forall $\delta t\le\Delta t$. Otherwise, the DW moves to the next metastable state after a time interval $T$, making a jump $S=\sum_{j=1}^L [n_j(t+T)-n_j(t)]$. Note that the jump is proportional to the average displacement of the DW between metastable states. To calculate the distributions of size $S$ and duration $T$ we extract the sequences of avalanches, over many realizations, between first- and last-jump, where the DW is finally pinned. 

\section{\label{sec:results}Results}

The distributions of avalanche-size $P(S)$ and av\-a\-lanche-duration $P(T)$ have been obtained for different values of $c$ and $p$ and different values os $L$ between $64$ and $4096$. Powers law-type behavior was  observed on both distributions.

\begin{figure}[th]
\includegraphics[width=.9\linewidth]{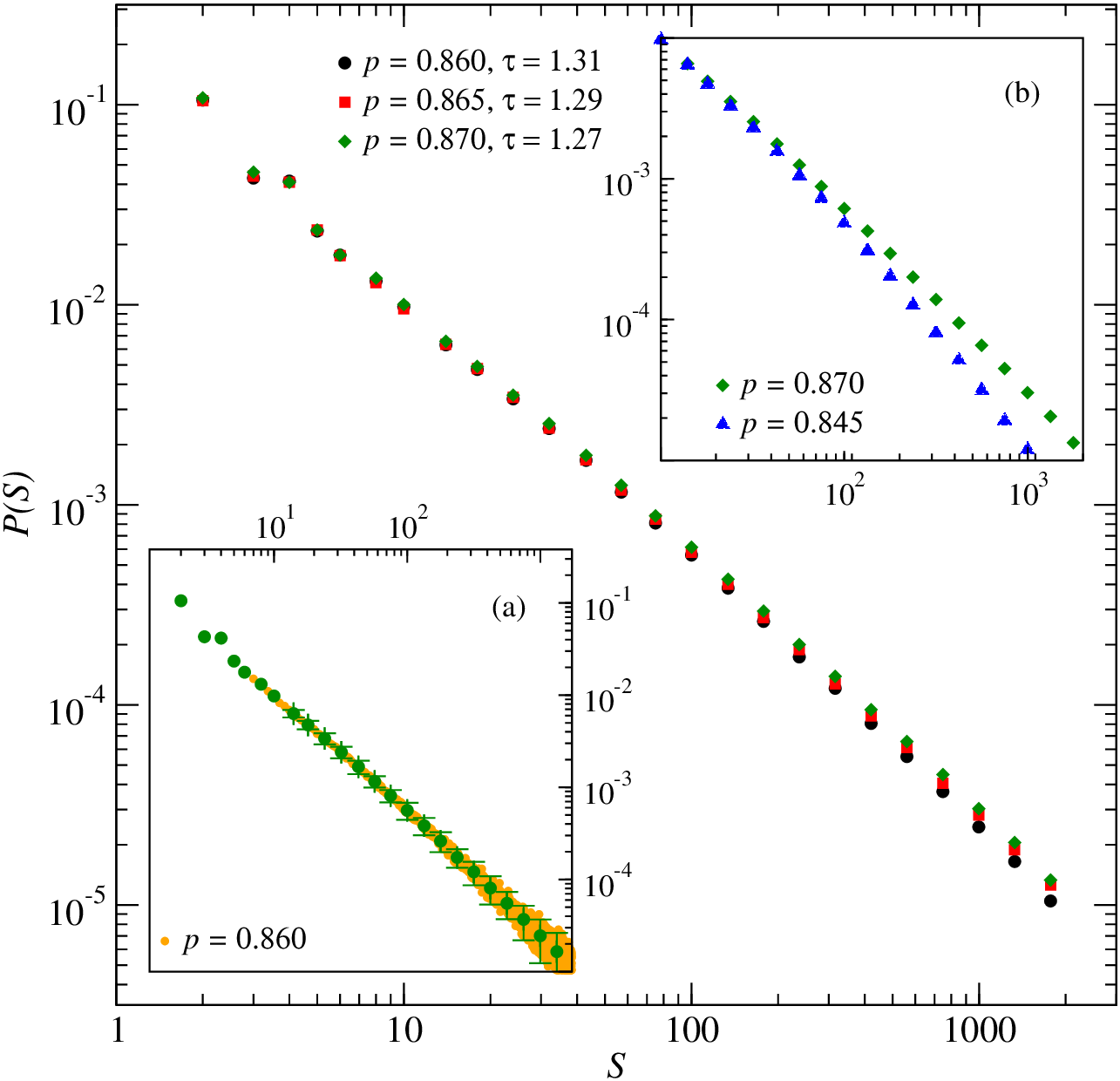}
\caption{(color online) Log-Log plot of normalized avalanche size distribution $P(S)$ with $L= 1024$, $c=10^{-3}$ and different values of $p$ [$0.860$ (circle), $0.865$ (box), $0.870$ (diamond)]. Inset: (a) Log-Log plot of the data as is were obtained ($p=0.860$) and the points considered for a better visualization logarithmically equispaced. (b) Log-Log plot of normalized avalanche size distribution with different values of $p$ [$0.870$ (diamond), $0.845$ (triangle up)].\label{one}}
\end{figure}

In Figure 1 we show  $P(S)$ for different densities of active sites $p$ close to criticality ($p_c=0.865\pm 0.005$) and $c$ fix close to zero. The critical exponent of avalanche size is $\tau = 1.29\pm 0.02$.
It is possible to note that as increases the value of $p$ diminishes the value of $\tau$. This behavior was observed for different values of $L$. The inset (b) shows that as decreases the value of $p$ the system moves away from the criticality, reducing the region that presents a behavior power law type.

In Figure 2 we show $P(S)$ for different values of $c$ and density of active sites fix $p=0.860$. In all the  cases, also for different sizes $L$ and density $p$ in the range $0.835-0.880$, it could observe a diminution of $\tau$ with the increase of c. For $c=0$ the system moves away from the power law type behavior dramatically as can observe on the inset (a), hence $c=0$ can be take as another critical point. From  inset (b), it is possible to notice that the size of the system does not affect considerably the behavior except for very high values of $S$ where finite size effect can be observed for the case of lattice with small edge $L$. 

Our results for $\tau$ are in excellent agreement with experimental results reported for ferromagnetic thin films. 
Because the determination of avalanche sizes in thin films is too difficult, the amount of experimental data is limited. Recently some data using magneto optics measurements was reported. For Co thin films, Kim {\sl et al.} reported power law behavior of the avalanche-size distribution, with critical exponent for various film thickness, with values between $\tau=1.34\pm 0.07$ (for 5 nm) and $\tau=1.30\pm 0.05$ (for 50 nm).\cite{Kim-03b} In a recent work Shin {\sl et al.} reported similar experimental results on Co films, with $\tau\simeq 1.33$ in the thickness range 5--50 nm.\cite{Shin-08} For MnAs films Ryu {\sl et al.} founded that the scaling exponent $\tau$ varies continuosly from $1.32\pm 0.06$ to $1.04\pm 0.05$ as increase the temperature from $20\,\celsius$ to $35\,\celsius$.\cite{Ryu-07} Magni {\sl et al.} performed multiscale measurements of avalanche distributions varying the observation window size on thin films of permalloy using a hight resolution Kerr magnetometer.\cite{Magni-09} They determined $\tau\simeq 1.18$ measuring at several window widths of an 170 nm film. More recently, Papanikolaou {\sl et al.} report measures of Barkhausen noise time series in 1 $\mu$m-thin films of permaloy using techniques inductive in an open magnetic circuit.\cite{*[{}][{. Supplementary Information on Films and Dimensionality at DOI: 10.1038/NPHYS1884}] Papanikolaou-11} According to our knowledge for ferromagnetic thin films, they first established the power law behavior for the distribution of avalanche duration with values for the exponent $\alpha$ in the mean-field universality class for 3-dimensional magnets.

\begin{figure}[h]
\includegraphics[width=.9\linewidth]{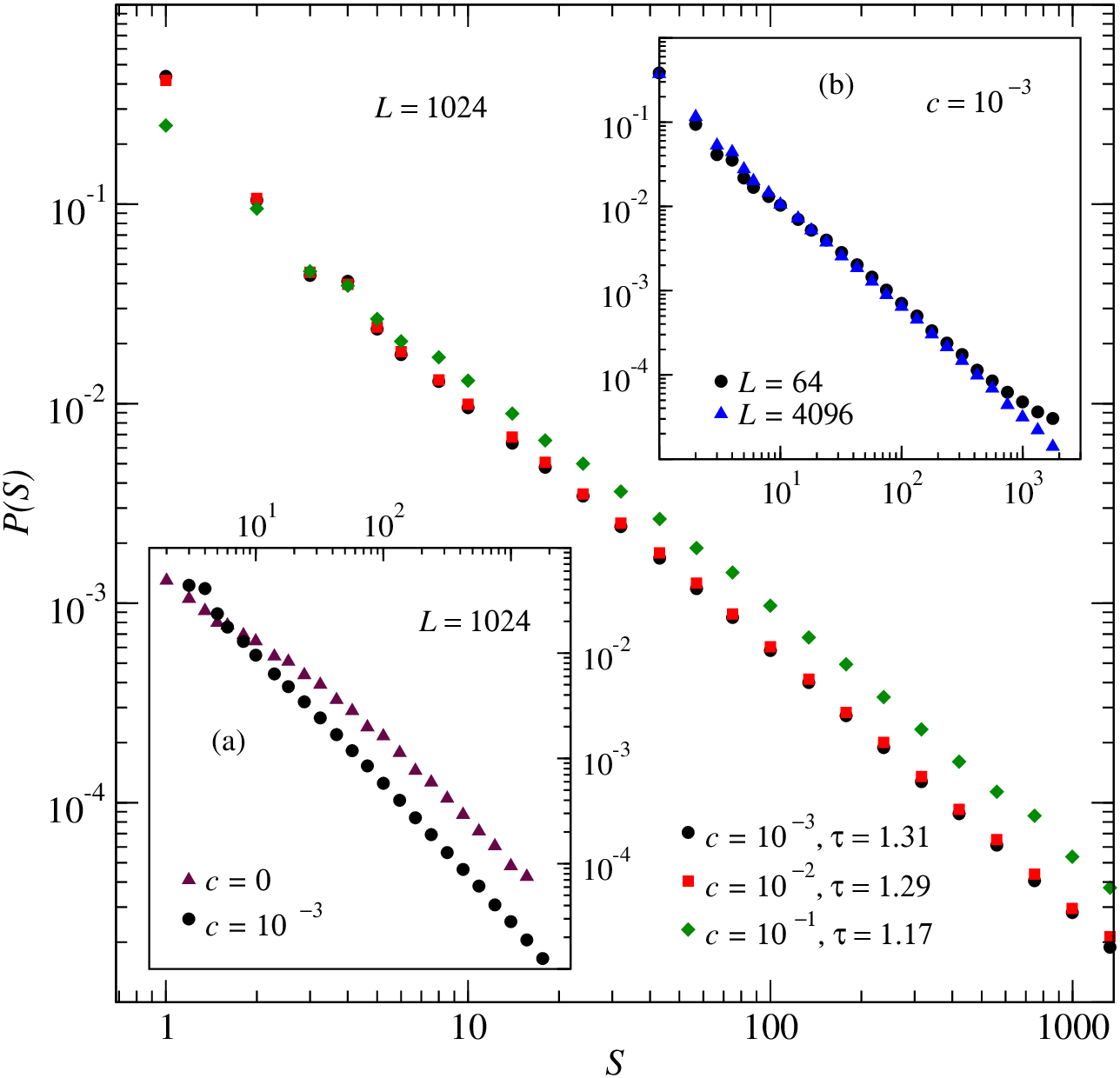}
\caption{(color online) Log-Log plot of normalized avalanche size distribution with $L= 1024$, $p=0.860$ and different values of $c$ [$10^{-3}$ (circle), $10^{-2}$ (box), $10^{-1}$(diamond)]. Inset: (a) Log-Log plot of normalized avalanche size distribution with $L= 1024$, $p=0.860$ and different values of $c$ [$10^{-3}$ (circle), 0 (triangle up)] (b) Log-Log plot of normalized avalanche size distribution with $c=10^{-3}$, $p=0.860$ and different values of $L$ [$64$ (circle), $4096$ (triangle up)].\label{two}}
\end{figure}

Rosso et al. recently calculated numerically the jumps distribution between successively pinned configurations in a 2-dimensional interface spring model of small intensity in a random-field landscape.\cite{Rosso-09} They study mostly random-field disorder (RF), but we also check that the results are the same for random-bond disorder (RB) and compare with predictions of the Functional Renormalization Group (FRG) field theory.\cite{Narayan-93, *Leschhorn-97} For RF disorder they determine $\tau=1.30\pm 0.02$ near to FRG prediction $\tau=1.2735\pm 0.0005$. Although that our value for $\tau$ is close, there are no other numerical or theoretical results for the exponent duration $\alpha$ for another comparison.  

Figure 3 shows $P(T)$ for $c=10^{-3}$, $p=0.865$ and different values of $L$. It is possible to note that on short duration jumps regime all the curves show the typical power law, with $\alpha=1.55\pm 0.05 $. This behavior is limited by the lattice size $L$. After certain jump duration $T^*$ for each $L$ the distribution presented a different behavior. This crossover from one regime to another could be attribute to finite size effect, wich is consistent with the fact that short (or large) duration jumps are possible in small (or giant) lattices. For $T>T^*$ the duration distribution $P(T)\propto\mathrm{e}^{- T/T_0}$ where was obtained from the simulation that $T_0\propto L^{\Delta}$ with $\Delta$ between 1.0 and 1.15. When $L\to\infty$ is interesting to note that $P(T)\to 0$ and $T^*\to\infty$, evidence of finite size effects. While the scaling regime remains essentially unchanged a peak occurs around the cutoff. The inset (b) shows in more detail the exponential regime where the dashed lines represents the fitting for each curve. 
%This behavior of finite size is observed for the plot of $P(S)$ in various publications.\cite{Queiroz-01,Durin-05}

\begin{figure}
\vspace{2ex}
\includegraphics[width=.9\linewidth]{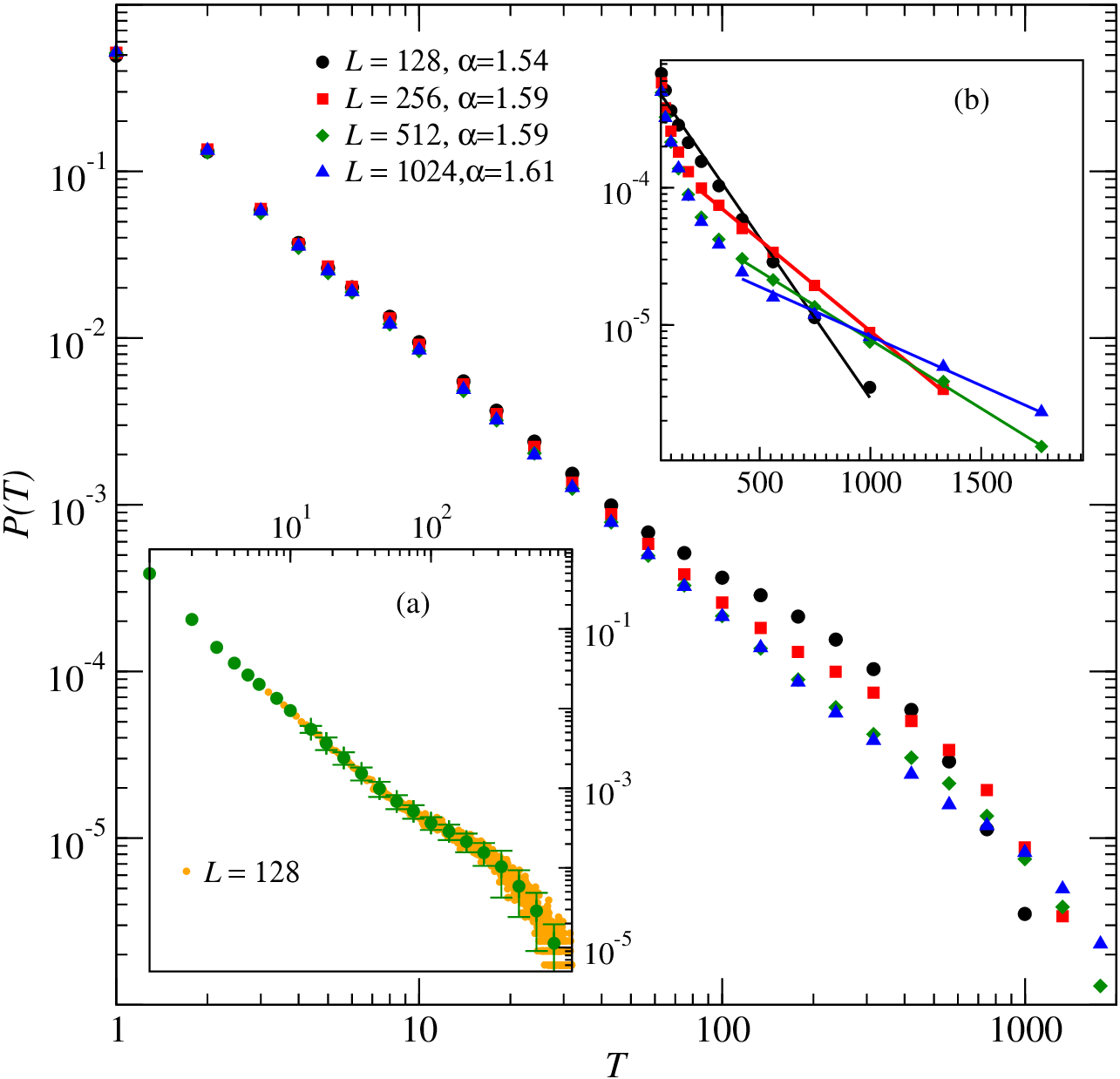}
\caption{(color online) Log-Log plot of normalized avalanche duration distribution with $c=10^{-3}$ and $p=0.865$ and different values of lattice edge $L$ between 128 and 1024. Inset: (a) Log-Log plot of the data as is were obtained ($L=128$) and the points considered for a better visualization logarithmically equispaced. (b) Semi-Log plot, detail of the central plot showing the effects of finite size, with an exponential fit in dotted line.\label{three}}
\end{figure}

\section{\label{sec:concl}Conclusions}

The main achievement of our work has been to introduce an avalanches model that describes the motion in disordered media of DW for magnetic thin film in which only short-range interactions are considered. The simulation results show the power law like critical behavior following the distributions of size and duration of Barkhausen jumps. From the results of the simulation, around the criticality, we observe that $\tau$ decreases when $p$ or $c$ increases.
At criticality the results for the exponent $\tau$ of the size distribution is in very good agreement with experimental and theoretical work scarce. On the contrary, as the exponent $\alpha$ of the duration distribution is not been measured experimentally for 2-dimensional magnets, our results can not be checked. Comparing with the experimental values for bulk magnetic materials where short-range interactions dominate, our values for $\tau$ are very close while those for $\alpha$ differ considerably. Our understanding of two-dimensional modeling for ferromagnetic thin-film remains incomplete. Other exponents should be calculated in order to establish the scaling relations, {\sl e.g.} exponent of growth and dynamic exponent, among others. In our model the avalanche size distribution does not depend significantly on the lattice size, since the power-law behavior extends over two logarithmic orders. By contrast, our model shows that the avalanches duration distribution includes finite size effects, which tend to disappear by increasing the lattice size according to an exponential cutoff. 
An outstanding question in thin films is the theoretical and experimental characterization of materials with predominantly long-range interactions.\cite{Mughal-10} 
Finally, models based on evolution rules could be ussefull to take into account long-range interactions.

\bibliography{bm-2011-physa}
\end{document}